# Tunable Optoelectronic Properties of WS$_2$ by Local Strain Engineering and Folding

Ahmed Raza Khan[1,3,§], Teng Lu[2,§], Wendi Ma[1], Yuerui Lu[1,*] and Yun Liu[2,*]

[1]Research School of Engineering, College of Engineering and Computer Science, Australian National University, Canberra ACT, 2601, Australia

[2]Research School of Chemistry, Australian National University, Canberra ACT, 2601, Australia

[3]Department of Industrial and Manufacturing Engineering, (Rachna College), University of Engineering and Technology, Lahore 54700, Pakistan

\* To whom correspondence should be addressed Yuerui Lu (yuerui.lu@anu.edu.au), Yun Liu (yun.liu@anu.edu.au),

[§]These authors contributed equally to the work.

**ABSTRACT.** Local-strain engineering is an exciting approach to tune the optoelectronic properties of materials. Two dimensional (2D) materials such as 2D transition metal dichalcogenides (TMDs) are particularly well-suited for this purpose because they have high flexibility and they can withstand high deformations before rupture. Local strain engineering in 2D TMDs is achieved via strained wrinkles. Wrinkles on thick TMDs' layers have been reported to show interesting photoluminescence enhancement due to bandgap modulation and funneling effect. However, the wrinkles in ultrathin TMDs have not been investigated because they can easily fall down to form folds in these ultrathin layers of TMDCs. Here, we have achieved both wrinkles and folds simultaneously in 1-3L WS$_2$ using a new fabrication technique. A layer dependent reduction in surface potential is found for both folded layers and perfect pack layers due to the dominant interlayer screening effect. Strain dependent modulation in semi conductive junction properties is observed for strain induced wrinkles through current scanning. Thermo-ionic modelling suggests that the strained (1.6%) wrinkles can lower the Schottky barrier height (SBH) by 20%. Upon illumination, SBH reduces significantly due to photo-generated carriers. Our results present an important advance towards controlling the optoelectronic properties of atomically thin $WS_2$ via strain engineering, with applications in optoelectronics, quantum optics and nanophotonics device fabrication.

**Introduction**

Tuning the optoelectronic properties of semiconductors is important for device fabrication and fundamental science.[1] Researchers have used various ways to tune the optoelectronic properties such as strain engineering,[2,3] doping,[4,5] defect engineering,[6] etc. Particularly, local strain engineering provides an effective and straightforward way to tune the optoelectronic properties of semiconductors such as bandgap, carrier mobility, binding energy of excitons,[2,3,7] which has therefore been broadly used to improve the device performance such as enhancement of solar cell efficiencies and generation of single photon emitters.[8–12]

Two-dimensional (2D) atomic crystals such as 2D transition metal dichalcogenides (TMDC) monolayers are an ideal platform to study strain engineering because they can withstand strain greater than 10%.[13] In addition, they can be folded or wrapped due to their high flexibility, making them promising candidates for stretchable and flexible electronics.[13–15] Using strain engineering of 2D TMDCs, researchers have successfully tuned optoelectronic properties.[16,17] For example, Desai *et. al.* showed considerable photoluminescence (PL) enhancement on atomically thin tungsten diselenide ($WSe_2$) upon strain application.[18] Strain effect is used to increase the career mobility of ultrathin $MoS_2$, thus leading to an improvement in the performance of $MoS_2$ based field effect transistor (FET).[19]

Local strain engineered in 2D TMDs is typically achieved through stretching of a flexible substrate.[20] This method involves the exfoliation of 2D material on a pre-stretched flexible substrate followed by the formation of strained wrinkles after the release of the stretched substrate.[3] The wrinkles in ultra-thin TMDs easily fall down to form folds due to their low bending rigidity.[20] Therefore, wrinkles formation in thick TMDs (layer ≥ 3) is only reported.[3,20] The excitons in thick wrinkles are reported to funnel towards the middle of the wrinkle's region; therefore, the wrinkles can change the local confinement potential of excitons.[3] Wrinkles are also reported to show exciting bandgap modification in atomically thin $MoS_2$[3].

Here, we have developed a new method, mechanical buckling of flexible substrate, to create both wrinkles and folds simultaneously in 1-3L $WS_2$ by successfully controlling the strain. The Kelvin probe force microscope (KPFM) and conductive atomic force microscope (CAFM) are used for surface potential mapping and current mapping, respectively. A layer dependent reduction in surface potential is found due to the dominant interlayer screening effect irrespective of the layers' orientation. The current mappings demonstrate strain tuning of semi conductive junction properties of strain induced wrinkles. This behavior is explained using Thermionic model which suggests 1.6% strain on wrinkle nanostructures reduce 20% of Schottky barrier height (SBH). Photo- Conductive Atomic force microscope (PCAFM investigation reveals that the SBH can be further lowered due to photo-enhanced current. Our results demonstrate an important advance towards controlling the properties of atomically thin $WS_2$ via strain induced wrinkling and folding. Moreover, these techniques could be applied to other 2D materials[21–30] and heterostructures[31–34] for applications in quantum optics, nanophotonics and optoelectronics.

## Results and Discussions

We employed mechanical buckling of the flexible substrate to achieve both wrinkles and folds simultaneously in 1-3L $WS_2$. The details of the fabrication method are shown in **Figure 1a**. (i) The $WS_2$ flakes are exfoliated on a pre-buckled Gel polymer film. (ii) After exfoliation, the buckled Gel film is released causing compressive forces on the exfoliated $WS_2$ flakes. The compressive forces generate big and tiny wrinkles. (iii) The big wrinkles are formed in the direction perpendicular to the compressive forces (x-direction) and they fall down to form folds, (iv) whereas the tiny wrinkles are generated in the direction parallel to the compressive forces (y-direction). The tiny wrinkles maintain their wrinkles like curvature which is explained later. In the text, the terms of big wrinkles and folds are used interchangeable. Without the specific notation, winkles present tiny wrinkles. A 3D schematic of the folds and wrinkles' formation after mechanical buckling is shown in **Figure 1b**. The buckled $WS_2$ sample is transferred onto a $Si/SiO_2$/gold electrode substrate.[35,36] Phase shifting interferometry (PSI)[22,24,37–39] is employed to identify the layer number of the $WS_2$ flakes **(Figure S1)**.

It is unambiguous that big wrinkles are formed perpendicular to the compressive forces (antiparallel red arrows) as depicted in optical image of buckled $WS_2$ **(Figure 1c).** Intriguingly, along with big wrinkles, some tiny wrinkles are also detected and their alignment is almost perpendicular to the big wrinkles as revealed by AFM topography (**Figure 1d**). The width of the tiny wrinkles (100-190 nm) is much smaller than that of the big wrinkles, which is the possible reason why there is no trace of tiny wrinkles under the optical microscope. The inserted plot (red rectangle) of **Figure 1d** gives detailed height profiles of both big and tiny wrinkles which indicate that the height of big wrinkles is much smaller than the tiny wrinkles. Moreover, height profile of tiny wrinkles presents a sharp peak-shaped cross-section in contrary to big wrinkles which present an almost rectangular shaped cross-section. According to this profile, we assume that big wrinkles form folds whereas tiny wrinkles are the wrinkles maintaining their curvature. **Fig. 1e** plots the height difference between the big/tiny wrinkle and corresponded flat regions. For the big wrinkles, the height differences measured on 1L, 2L and 3L samples are 1.4 ± 0.5, 2.8 ± 0.5 and 4.2 ± 1 nm, respectively. These values match the height of 2L, 4L and 6L samples very well as the thickness of single layer is evaluated around 0.7 nm, which agrees well with the reported value.[40] Therefore, the big wrinkles can be regarded as the bi-folded samples. *E.g.*, the big wrinkle on the 1L sample can be considered as bi-folded single layer (1L+1L+1L) samples and its height is in good consistency with the pristine 3L sample. In a previous work on $MoS_2^3$, the appearance of the fold regions is possibly ascribed to the collapsing of the large wrinkles caused by the buckling-induced delamination. Such wrinkle to fold transition is also observed on monolayers of graphene and TMDC by researchers in the past.[20,41–43] It is worthy to note that the tiny wrinkles appear in the center of the nano-flakes and therefore, their height values present large deviations. However, their deformation is relatively small and they maintain their shape like a curvature. Wrinkling requires a minimum value of strain to initiate the wrinkles formation. As the strain increases, the height of the wrinkles increases. After a critical strain level ($\varepsilon_f$), wrinkles cannot maintain their curvature and collapse forming folds **(Figure S2c)**. Critical strain level is dependent on layer thickness and elastic modulus of ultra-thin $WS_2$ film (See section S3 for more details). As layer number increases, $\varepsilon_f$ increases and fold formation in higher layer number decreases. Therefore, wrinkle to fold transition is dominant in small layer number (1L > 2L > 3L) due to their low critical strain ($\varepsilon_f$). Moreover, it can also be found that wrinkles' average height increases as layer number increases (3L > 2L > 1L) which can be attributed to increasing critical strain level ($\varepsilon_f$) in higher layer number showing more capability to maintain wrinkles' curvature **(Figure 1e)**. Additionally, the width of the folds also showed an increase as layer number increases which ranges 180-400 nm, 250-450 nm and 350-500 nm for 1L, 2L and 3L respectively.

**Surface Potential Measurements.** With intentional manufacturing, we have obtained different local features (flat, fold and wrinkles) in 1-3L $WS_2$. As varying the local stacking and strain states of the WS$_2$ samples, the electronic structures are expected to be tuned. KPFM is a useful tool to map the surface potential at the nano-meter scale. In KPFM, an AC bias is applied to a conductive tip to produce an electrical force which is minimized when contact potential difference or surface potential between tip and sample is compensated by an applied DC bias. In principle, the contact potential difference ($V_{CPD}$) is originated from the different work functions between the tip and sample, which can be expressed as follows;[44,45]

$$V_{CPD} = \frac{\Phi_{tip} - \Phi_{sample}}{-e} \qquad (1)$$

Where $\Phi_{tip}$ and $\Phi_{sample}$ are work functions of the tip and the sample and $e$ is the electronic charge. However, in reality, many factors such as trapped charges, tip damage and permanent dipoles between the sample and tip will impact the results. Therefore, KPFM usually gives an accurate measurement of the potential difference rather than the absolute value. The surface potential mapping is carried out for the flat, folded, wrinkled regions of 1-3L WS$_2$. The typical morphology and surface potential images are presented in **Figure 2a-2d.** The potential measured on different layers presents evident contrast. Combining the morphology mapping, in the 1L, 2L and 3L samples, the surface potential in fold regions is unambiguously lower than the flat regions for the similar layer number. However, no clear features are obtained for wrinkles in surface potential mapping which can be ascribed to the small size (width~100-190nm) of wrinkles making it to be difficult to be resolved. In order to conduct quantitative analysis on folded regions, the surface potential values are calculated by fitting the histograms collected from different domains **(Figure S3).** The offsets (**Figure 2b** and **2d**) exhibit a small divergence, therefore, the surface potential of flat 3L sample is considered as the reference to make the measured values comparable. As shown in **Figures 2e** and **2f**, the surface potential of flat and folded regions decreases linearly with an increase in layer number. The differences between each layer are ~23 mV and 9.5 mV for flat and folded regions, respectively. Taking the KPFM set-up into the consideration, work function increases as layer number increases which is consistent with the previous studies on layer dependent work function of flat $MoS_2$.[46,47] The layer-dependent work function can be explained by the dominant interlayer screening effect reported for both graphene and $MoS_2$.[48–50] Geometrically, the layer number of folded regions is comparable with the pristine 3L, 6L and 9L samples while the twisted stacking will result in the change of the interlayer coupling in comparison with the perfect packing samples. Intriguingly, no distinct difference is obtained between the surface potential of bi-folded 1L and flat 3L samples, indicating the interlayer screening effect is dominant in $WS_2$ folds irrespective of the layer orientation.

**Dark carrier transport.** The charge transport behavior is important characteristic for the semi conductive device performance. This section will focus on charge transportation for flat, folded and winkled regions. A schematic of the set-up for the conductive and photoconductive atomic force microscopy (CAFM and PCAFM) is shown in **Figure 3a**. The CAFM tip acts as the top electrode and gold is the bottom electrode. In dark condition (no light), the current mappings of AFM topographic image (red rectangle) are done under 4V and 10V bias (**Figure 3b-3c**). No evident current is detected overall under 4V except some traces at the wrinkled regions. With increasing the bias to 10V, the wrinkles exhibit significant enhancement of electrical current flow and current reaches around 400 pA. The local *I-V* curves on different regions are measured to further investigate the electronic properties as shown in **Figure 3d**. At the low bias, the current measured at flat, folded and wrinkled regions exhibit nearly insulating behavior while the diode-like characteristics are obtained by gradually increasing the bias. Schottky barrier height (SBH),

$\Phi_b$, measured at wrinkled regions is lower than those measured at flat and folded regions which will be explained later. The distance to the bottom gold electrode is similar (~20 μm) for three regions. Therefore, we can assume the obtained *I-V* behaviors are mainly determined by the metal tip-semiconductor contact resistance. Due to the difference between the tip work function and electron affinity of WS$_2$, Schottky barrier is formed at the interface. **Figure 3e** illustrates the equilibrium case of Schottky barrier between CAFM tip and WS$_2$ (flat and wrinkle). It is evident SBH represented by $\Phi_b$ ($\Phi_b = \Phi_{tip} - \chi$, where $\chi$ is the WS$_2$ electron affinity), is the determinant factor for the current flow.[51,52] Li et al.[49] performed a C-AFM study on ultrathin MoS$_2$ using a conductive tip and successfully explained the characteristics of the resulting junction in the forward bias regime using thermionic emission model. Therefore, the current under forward bias can be described by the thermionic emission model over the Schottky barrier as under [53];

$$I = A_{tip} A^* T^2 e^{-\frac{q\Phi_B}{kT}} e^{\frac{q(V-IR)}{nkt}} \qquad (2)$$

where $A_{tip}$ is the tip-sample contact area, $A^* = 4\pi q k^2 m_{eff}/h^3$ is the Richardson constant with effect mass for electrons ~ $0.3 m_e$ for WS$_2$ [54], $h$ is the Planck constant, $T$ is the temperature which equals ~ 300 K, $q$ is the elementary charge and $\Phi_B$ is the SBH, $k$ is Boltzmann constant, $R$ is the resistance in the circuit, and $n$ is the ideality factor. Since the current value is quite small (< 500 pA), the voltage drops across the Schottky barrier (*V-IR*) is almost equal to the applied bias. The fitting parameters for the three curves are displayed in **Figure 3d**. The large ideality factor, $n$ = 34, 35 and 41 for flat, fold and wrinkle suggests the thermionic emission model requires careful consideration while $\Phi_b$ fitted for the flat sample is around 603 meV, which is close to the recently reported $\Phi_b$ =590 meV between the Pt-coated tip and WS$_2$ nanoflake.[55] The fitting results suggest different *I-V* characteristics collected from flat, folded and wrinkled regions are originated from the different SBH formation at the metal-semiconductor interface. A large reduction of $\Phi_b$ (~486 meV) is obtained for the wrinkled WS$_2$ junction as compared to the flat WS$_2$ junction whereas the strain is calculated to be 1.6% on the wrinkle using equation (1) in the supplementary information. Previously, researchers have reported strain induced bandgap reduction in ultrathin WS$_2$[7,56] which can be attributed to lowering of SBH in wrinkled WS$_2$. The bandgap generally decreases with increasing the thickness in 2D TMDs semiconductors.[57,58] Therefore, we find a reduction in SBH of tip-WS$_2$ (fold) junction which is in good agreement with a previous study showing SBH reduction with increasing the thickness of MoS$_2$ in MoS$_2$-metal contact.[59] However, it is still worthy to note that the folding of 2D materials could change the stacking orders among the layers, thereby tuning interlayer coupling and band structures.[60].

**Photo carrier transport.** In the previous sections, we successfully demonstrated strain engineering and folding as powerful tools to tune the electronic properties. Strain engineering and folding are also expected to tune optoelectronic properties. In this context, the current mappings of the topographic image (red rectangle in **Figure 3b)** are carried out under dark (no light) and light condition for direct comparison. It is evident that current flow under illumination is enhanced for all the regions (flat, fold and wrinkle). However, wrinkled regions present the strongest contrast compared with other two regions. *I-V* junction characteristics of flat, folded and wrinkled region under illumination are presented to further explore this behavior **(Figure 4b)**. Here, *I-V* curves present rectifying characteristics with 5V forward bias under illumination whereas no obvious current responses were obtained for three regions using the same applied voltage under the dark condition. Furthermore, the photocurrent is not obvious under the zero, which might be arisen from the tiny photo-response compared with the instrumental current threshold (~10 pA) or small work function difference between the top and bottom electrodes

(Pt/Ir tip and Au). Enhancement of the electrical curent under illumination might have a close relationship with the photo-generated carriers. As the *I-V* curves still exhibit diode-like features, it can be proposed that the current flowing in the circuit is mainly decided by the metal/semiconductor contact resistance. Through adopting the thermionic model, either increasing the temperature or decreasing the SBH can intensify the current. Both current mapping and time-resolved photo-response under the repeated on-off light irradiation indicate the hysteresis, which is normally due to the thermal effect, is not obvious in flat $WS_2$ samples (**Figure S5**). To further investigate the impacts of photo-generated carriers on SBH, the surface potential under the light irradiation is shown in **Figure 4c**. It is obvious that the surface potentials on both flat and folded regions exhibit an abrupt increase when the incident laser is on. **Figure 4d** shows the surface potential variation as a function of the incident laser power measured in the same region. The detailed surface potential images can be found in **Figure S6**. For both the flat and folded regions, the biggest variation occurs when the laser is switched on, which indicates that the surface potential change mainly originates from the photo-excitation rather than the thermal-excitation. According to the KPFM set-up, the increase of surface potential is a consequence of accumulated positive charges on the surface or the appearance of upward dipole-moment.[61] Both effects can bend the band structure of $WS_2$ samples, lowering the SBH. Therefore, the photo-enhanced current is attributed to lowering the SBH by the photo-generated current.

## Conclusion

In summary, we have successfully demonstrated the fabrication of both fold and wrinkle nanostructures simultaneously in 1-3L $WS_2$ by controlling strain. It is found that the interlayer screening effect is the dominant factor in layer dependent surface potential measurements.

Therefore, we find layer dependent surface potential reduction for both perfect pack and twisted layers of ultrathin WS$_2$. The current mappings demonstrate strain tuning of semi conductive junction properties of strain induced wrinkles. This behavior is explained using Thermoionic model, which suggests 20% reduction in Schottky barrier height through 1.6% strain on wrinkle. Photo- Conductive Atomic force microscope (PCAFM) investigation reveals further lowering of SBH due to photo generated carriers. Our technique offers a route to local strain engineering in ultra-thin materials, opening up many applications in diverse fields such as electronics, quantum optics, optoelectronics and surface science.

**Experimental section**

**Sample Fabrication.** WS$_2$ flakes were exfoliated onto buckled elastomeric substrate (Gel-Film® WF 6.0mil ×4 films) using scotch tape. Subsequently, the Gel film is suddenly released, generating well-aligned folds and wrinkles in WS$_2$ layers due to the application of compressive forces because of the sudden release. Sudden release of the pre-stress films was found to give a higher yield of folds and wrinkles in WS$_2$ layers. Strained WS$_2$ sample was transferred onto Au/SiO$_2$/Si electrode substrates followed by adhesively bonding an iron pad using copper tape to connect with AFM for further characterizations. AFM, PSI and Raman spectroscopy were used to detect the layer numbers.

**Surface Potential Measurements.** Surface potential measurements were done using AFM (Asylum Research, Cypheras) after carrying out the procedures described in the Sample Fabrication section. Pt/Ir coated Si tip (nanosensor PPP-EFM) with a calibrated spring constant ~ 1.9 N/m and radius of 28 ± 10 nm was used to conduct KFPM measurement. The tip was scan above ~ 10 nm higher than the surface in the noncontact mode with a drive frequency of 70 kHz and 1 V AC voltage. The SP mapping images by KPFM (scan size: 3μm ×3μm) were obtained, where the temperature was maintained at room temperature.

**Dark and Photo-current Scanning.** The Current and Photocurrent measurements were conducted on AFM in ambient conditions using Pt. tip, as illustrated in Figure 3a. Contact mode was used during the current scanning AFM. Topographic and Current images were obtained simultaneously so that topography and local currents can be compared directly. Most images were 512×512 pixels. For photo-current measurements, the illumination source was 532 nm laser. During certain measurements, neutral density filters were used to modulate the laser intensity. AFM images were analyzed and plotted using the Gwyddion software package.

**Conflict of Interest**. The authors declare no competing financial interest.

**Acknowledgments**. We would like to acknowledge the financial support from Australian National University, Australia and Faculty Development Program funding under University of Engineering and Technology (UET) Lahore, Pakistan and Rachna College of Engineering and Technology (RCET), Gujranwala, Pakistan.

**Supporting Information Available**. All additional data and supporting information are presented in the supplementary information file.

**Figure Caption list**

**Figure 1 | Creation of wrinkles and folds in 1-3L WS$_2$. (a)** Schematic diagram of the fabrication process for wrinkles and folds formation in atomically thin WS$_2$. (i) An elastomeric substrate (Gel film) is buckled prior depositing WS$_2$ flakes using a tape. (ii) The buckled substrate is released causing compressive forces which create big and tiny wrinkles perpendicular and parallel respectively, to the direction of forces (X-direction). (iii) Big wrinkles fall down to form folds (X-direction view) (iv) whereas tiny wrinkles retain their shape (Y-direction view). **(b)** A 3D image of strained sample showing wrinkles and folds, different colours for the wrinkles and the folds are used for clear visibility. **(c)** Optical microscopic image of 1-3L strained WS$_2$ sample fabricated by the method described above; folds are created perpendicular to the direction of the compressive forces, two anti-parallel arrows (red colour) indicate the direction of the compressive force.**(d)** AFM topography image of the region marked by the white dashed rectangle in (c), AFM measurement of the wrinkle (20 nm height and 120nm width) and fold (2.8 nm height and 450nm width) on 2L are in green and blue colours respectively shown in zoom-in image of the region marked by the red rectangle. **(e)** The layer dependence of wrinkles and height differences for folds measured on 1L, 2L and 3L. The histogram shows the AFM height measurements, with variation in measurements indicated by the error bars. The orange and green rectangles indicate the height measurements for folds and wrinkles respectively.

**Figure 2 | Layer-dependent surface potential of folds in 1-3L WS$_2$.(a)** AFM topography image, showing the flat, folded and wrinkled regions on 1L and 3L.**(b)** Surface potential mapping of (a). **(c)** AFM topography image, showing the flat, folded and wrinkled regions on 2L and 3L. **(d)** Surface potential image of the region shown in (c) **(e)** Surface potential histogram of 1-3L flat WS$_2$ (upper panel) and 1-3L fold WS$_2$ (lower panel). **(f)** Surface potential of 1-3Lflat WS$_2$ (red) and 1-3L fold WS$_2$ (blue), with variation in the readings is indicated by the error bars.

**Figure 3 | Current mapping of wrinkles in atomically thin WS$_2$ by conductive AFM imaging. (a)** A schematic of conductive AFM configuration for carrier transport measurements. Strained sample is transferred onto a gold electrode and the voltage is applied on the sample under forward bias. **(b)** AFM topography image, showing the flat, folded and wrinkled regions. **(c)** Current mapping of the region marked by red rectangle in (b) under bias voltages of 4V and 10V, **(d)** A current-voltage (*I-V*) curve for flat, fold and wrinkle indicated by red dots in (b) under (-10 to +10V) bias voltage, Schottky barrier height ($\Phi_b$) for flat = 603 meV, fold = 566 meV and wrinkle = 486 meV is calculated using thermionic model. The corresponding strain for the wrinkle is calculated to be 1.6%. **(e)** Energy band diagram for CAFM tip-WS$_2$ (flat) and CAFM tip-WS$_2$ (wrinkled) junction. $\Phi_{tip}$ is the CAFM tip work function, $E_{vac}$ is the reference vacuum level, q is the charge, $\Phi_b$ is the barrier height, $\chi_f$ and $\chi_w$ are electron affinity for flat and wrinkle WS$_2$ respectively, $E_{ft}$, $E_{ff}$ and $E_{fw}$ are the Fermi level of the tip, flat and wrinkle WS$_2$ respectively. $E_{vf}$ and $E_{cf}$ are valence and conduction band levels of flat WS$_2$ whereas $E_{vw}$ and $E_{cw}$ are valence and conduction band levels of wrinkle WS$_2$.

**Figure 4 | Differentiation of wrinkled and folded nano-structures by photo-conductive AFM. (a)** Current mapping (of the region marked by red rectangle in Figure 3c) under dark (no light) and light (60 mW) at forward biased voltage of 4V; **(b)** Current-voltage (*I-V*) curve showing photocurrent response for flat, fold and wrinkle under excitation from a 532nm laser with 60 mW power density. **(c)** Surface potential mapping of 1L and 3L under dark (no light) and light (60 mW) **(d)** Laser power dependent surface potential (mV) response of 1 & 3L flat and fold under excitation from a 532nm laser with 60 mW power density. Error bars indicate the variation in measurements.

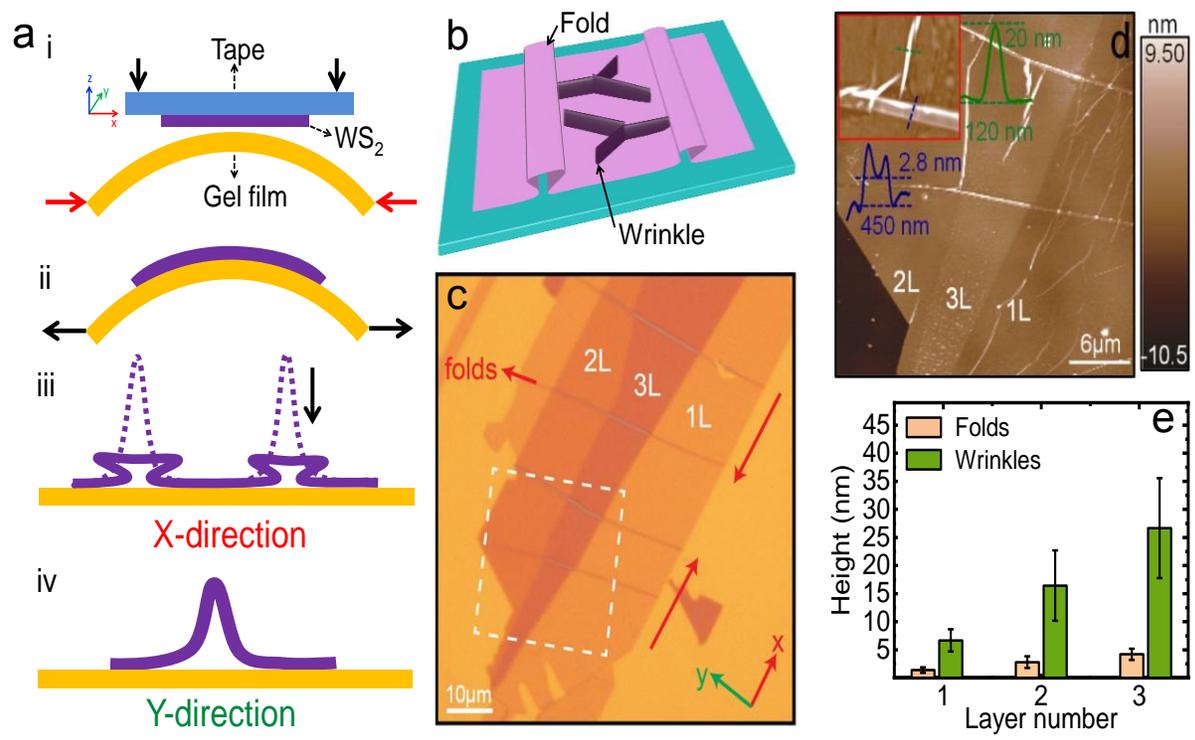

**Figure 1**

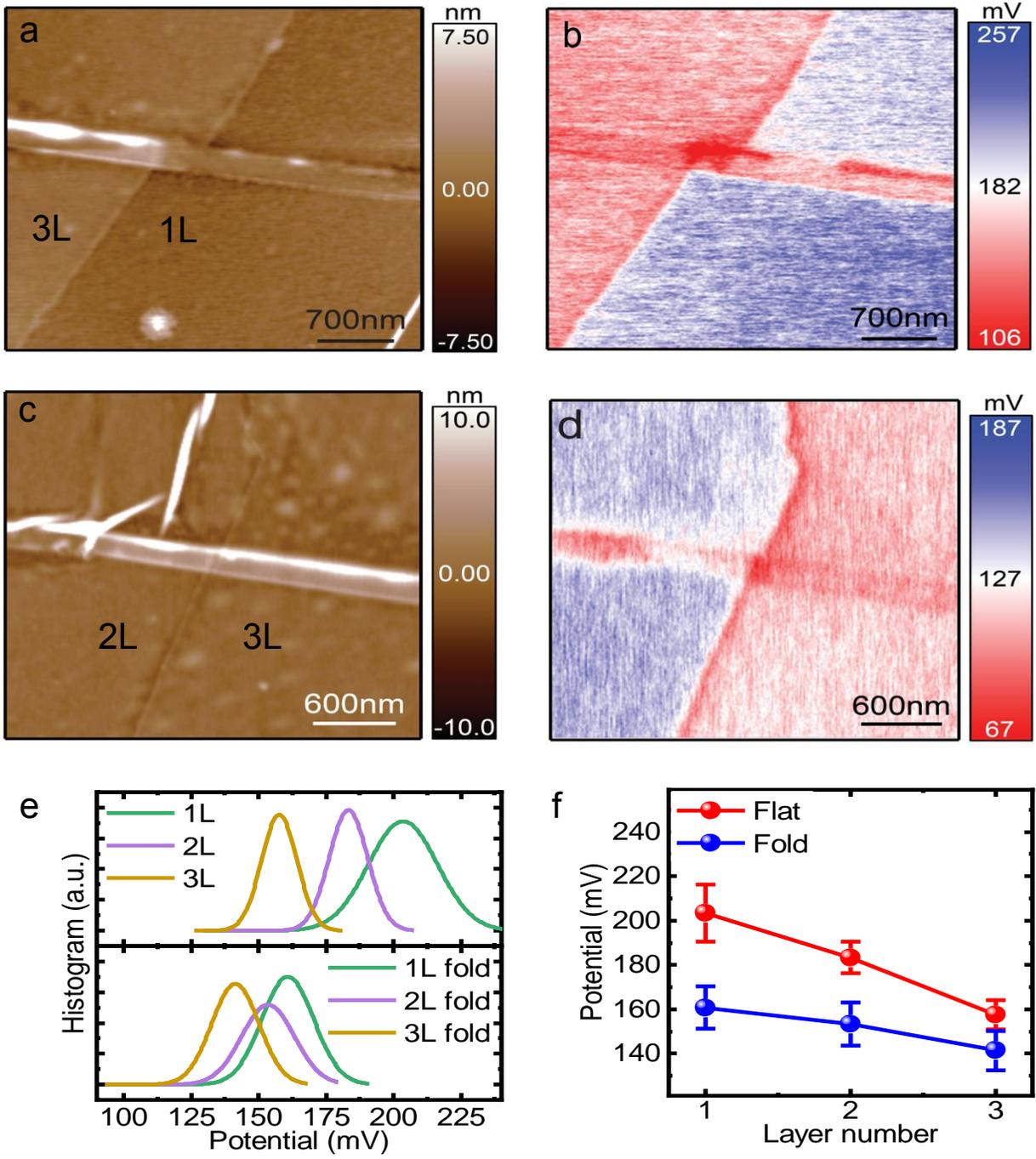

Figure 2

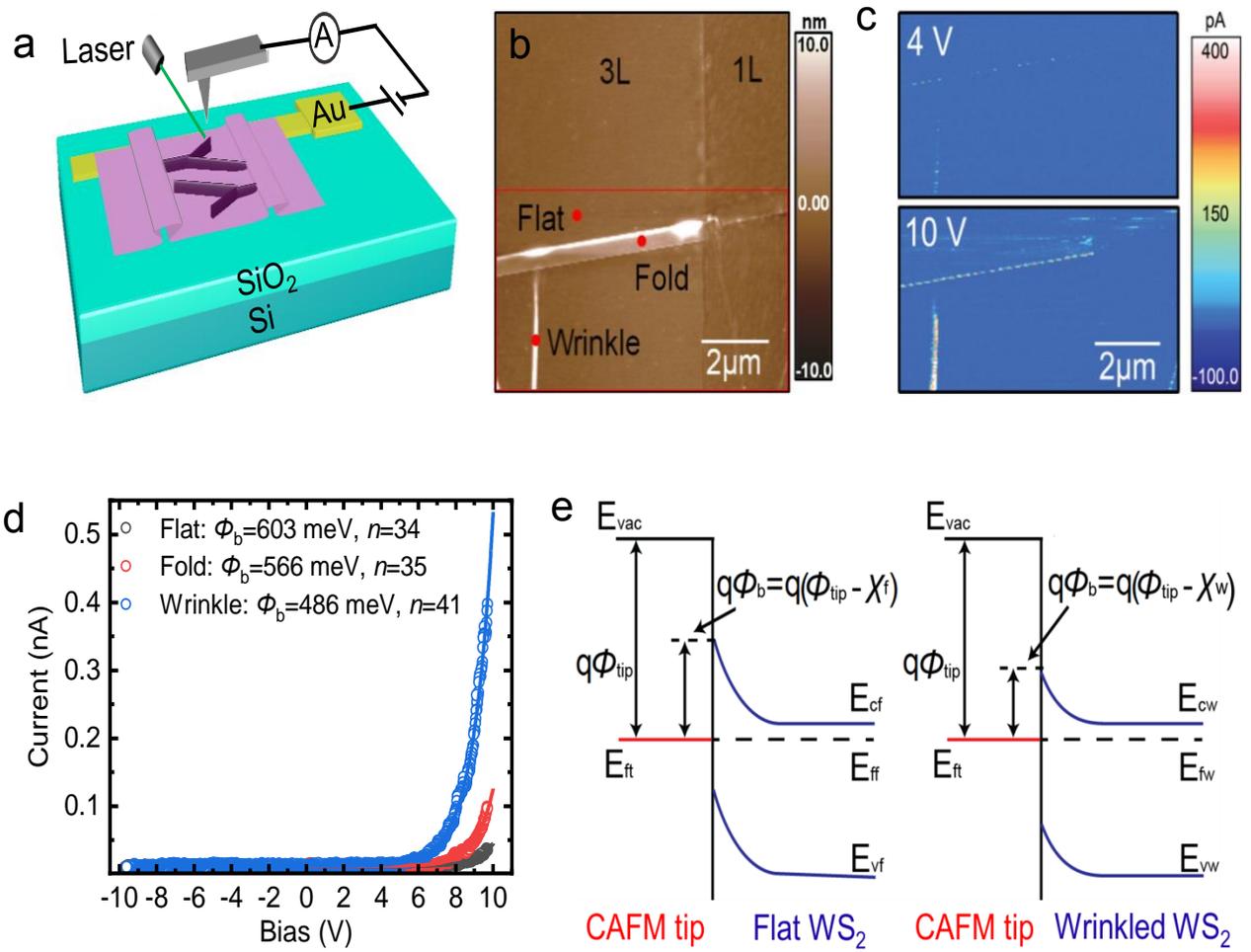

**Figure 3**

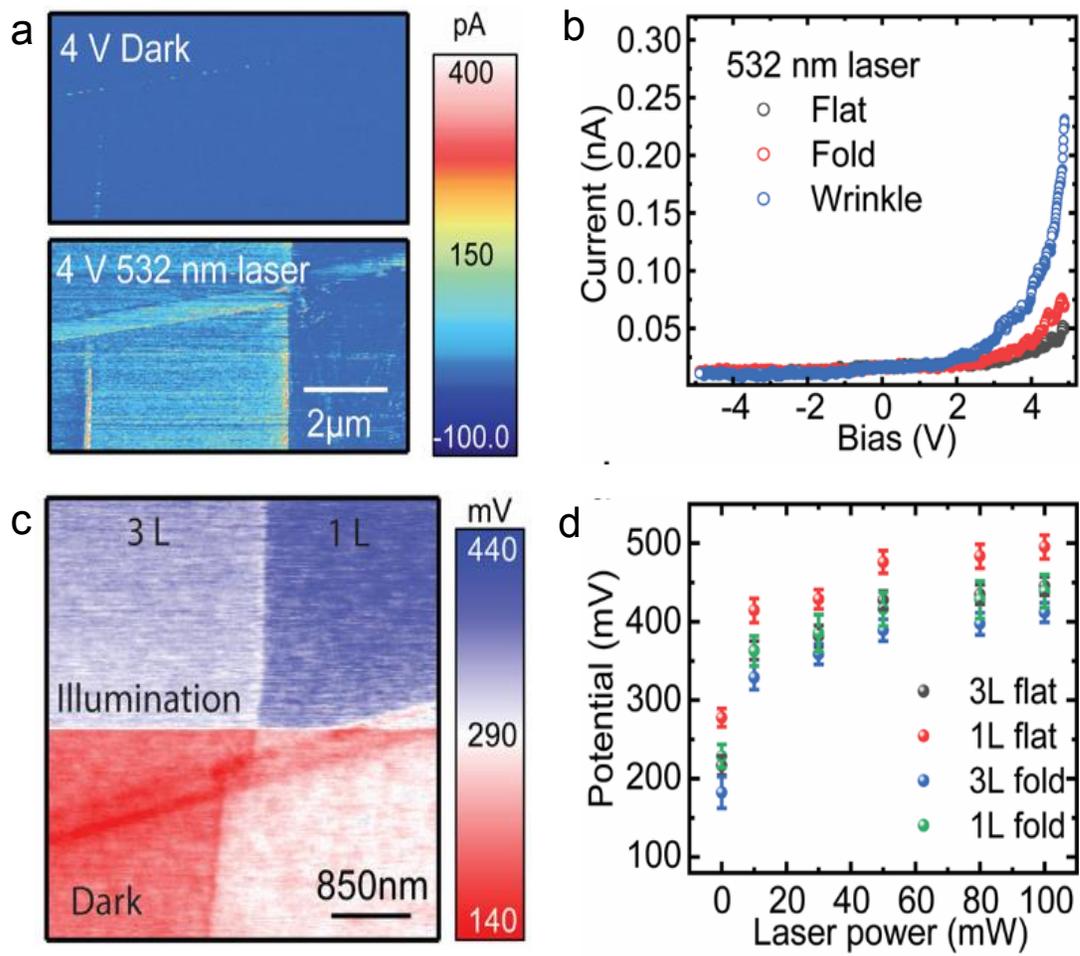

**Figure 4**

# Supplementary Information

# Tunable Optoelectronic Properties of WS$_2$ by Local Strain Engineering and Folding

Ahmed Raza Khan[1,3§], Teng Lu[2§], Wendi Ma[1], Yuerui Lu[1*] and Yun Liu[2*]

[1]Research School of Engineering, College of Engineering and Computer Science, Australian National University, Canberra ACT, 2601, Australia

[2]Research School of Chemistry, Australian National University, Canberra ACT, 2601, Australia

[3]Department of Industrial and Manufacturing Engineering, (Rachna College), University of Engineering and Technology, Lahore 54700, Pakistan

**\*** To whom correspondence should be addressed: Yuerui Lu (yuerui.lu@anu.edu.au) Yun Liu (yun.liu@anu.edu.au),

[§] These authors contributed equally to the work.

## S1. Strained Sample Fabrication

Strained $WS_2$ flakes by exfoliating $WS_2$ nano-layers are fabricated onto buckled elastomeric substrate (Gel-Film® WF 6.0mil ×4 films) using scotch tape. Gel-Film® is Polyester based commercially available elastomeric film. Subsequently, the Gel film is released to generate compressive forces on $WS_2$ nano-layers as sudden release causes compression force generating well-aligned folds and wrinkles in $WS_2$ layers. It is found that sudden release of the pre-stress films gives a higher yield of folds and wrinkle nanostructures within $WS_2$ flakes. The mechanism behind the formation of these folds and wrinkles is buckling-induced delamination. Such strained $WS_2$ samples were then transferred onto $Si/SiO_2$/Gold electrode substrates for further characterization.

## S2. Layer number identification of WS$_2$ on Si/SiO$_2$ substrate.

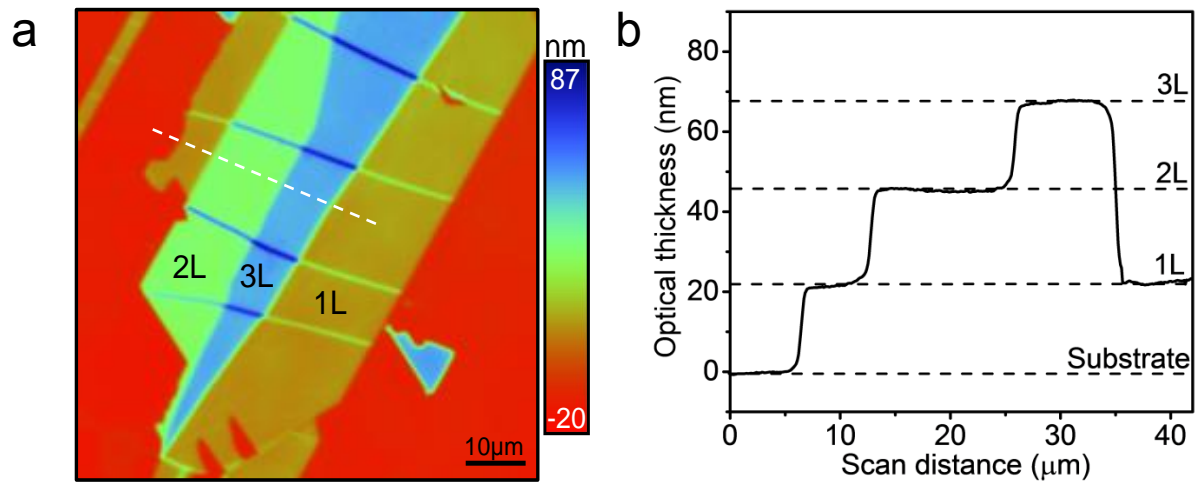

**Figure S1 | Layer number identification of WS$_2$ on Si/SiO$_2$ substrate.** (a) PSI (Phase Shifting Interferometer) image of WS$_2$ flakes showing 1L, 2L and 3L. (b) PSI-measured *optical path length* (*OPL*) values showing layer number for 1-3L WS$_2$ along the white dashed line in (b) using the methods described in [1,2,3]

## S3. Fold formation in 1-3L WS$_2$

This section describes the fold formation in 1-3L WS$_2$. Investigating the strained WS$_2$ sample, it is observed that the edges before and after big wrinkles (perpendicular to the direction of compressive forces) are not collinear (**Figure S2a**), instead are displaced by few tens of nanometers which showed the wrinkle to fold transition in ultra-thin WS$_2$.

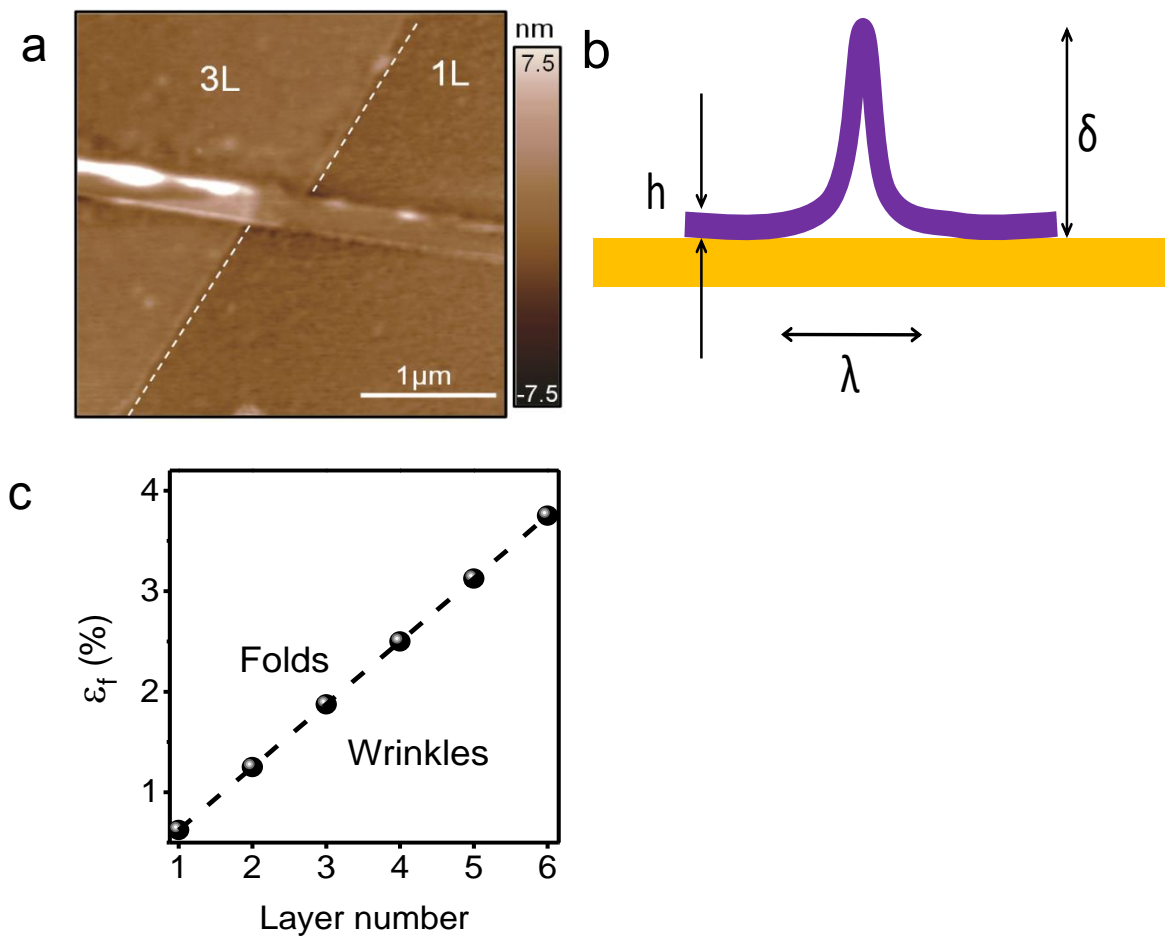

**Figure S2 | Fold formation in 1-3L WS$_2$.** (a) AFM topography image of a strained WS$_2$. The white dashed lines indicate that the edges before and after the fold are not collinear, thus, showing the collapse of the wrinkles during the buckling-induced delamination process. (b) Analysis of the

strain required for fold formation, $h$, $\lambda$ and $\delta$ are thickness, width and height of WS$_2$ sheet. (c) Layer dependent folding strain ($\varepsilon_f$) calculation for 1-6L. Wrinkles maintain their curvature before $\varepsilon_{f(\%)}$ and become fold after $\varepsilon_{f(\%)}$

We found that big wrinkles cannot maintain their wrinkles' curvature and tend to collapse forming folds,[4] and therefore, their height is found to be much smaller than tiny wrinkles. The total height measured on 1L, 2L and 3L samples match the height of 3L, 6L and 9L. We found tiny wrinkles' (formed parallel to the direction of compressive forces) height is greater than folds height, showing that they maintain their wrinkles' shape. The minimum strain required to start the wrinkling process is called the critical strain for wrinkling[5] The wrinkles height and density of wrinkles increases as the applied strain increases.[6–8] Eventually, after a strain level ($\varepsilon_f$), wrinkles cannot maintain its curvature and collapse. The opposite layers of wrinkles join each other to form folds. The folding process is dominant in small layer numbers which is greater in 1L than the 2L and 3L (1L > 2L > 3L) due to smaller value of elastic modulus in 1L as compared to 2L and 3L. As the layer number increases, critical strain for folding ($\varepsilon_f$) increases and fold formation becomes almost absent in higher number layers (5L and higher) due to high bending rigidity [9,10]. Here, we present the theoretical framework that we employ to find the layer dependent effect on the strain required for folding for WS$_2$.

The maximum uniaxial strain $\varepsilon$ is accumulated on top of the winkles and can be estimated as [9]

$$\varepsilon \sim \pi^2 h\delta/(1-v^2)\lambda^2 \qquad (1)$$

where $v$ is the Poisson's ratio ($v$ =2.2 for WS$_2$[11]), $h$ is the thickness of the flake, and $\delta$ and $\lambda^2$ are the height and width of the wrinkle which were measured using atomic force microscopy (AFM). Yuri et. al.[5] approximated the critical strain required for wrinkle to fold transition in thin films as;

$$\varepsilon_f \sim \left(\frac{h}{L}\right)\left(\frac{E_f}{3E_s}\right)^{\frac{1}{3}} \qquad (2)$$

where $h$ is the thickness of the flake, $E_f$ is the elastic modulus of the film (WS$_2$) and $E_s$ is the elastic modulus of the substrate (Gel film). For $E_f$(WS2), we use the relation as under;

Elastic Modulus = $E_{2D}/h$ [12], where $E_{2D}$ is 149 N/m[13] for 1L WS$_2$ and $h$=0.7nm for 1L give us $E_f$ (WS$_2$) = 198.6 GPa, As Gel-film is a polyester based elastomer material and $E_s$ = 0.92 GPa.[14] L is the sample length to define the boundary of the region where folding is expected. In order to compare the folding strain among layer numbers, we calculate $\varepsilon_f$ by taking L as the length of unit wrinkle along the direction of the force or average width of a wrinkle (500nm). Figure S3 shows layer dependent folding strain ($\varepsilon_f$) calculation for 1-6L. Wrinkles maintain their curvature before $\varepsilon_{f(\%)}$ and become fold after $\varepsilon_{f(\%)}$. Taking an average 1L tiny wrinkle size with 150 nm width, 7 nm height gives us an approximated 0.05% which is below than 0.63% which is calculated critical folding strain for 1L. As layer number increases, folding strain increases linearly (**Figure S3**), therefore, we experience wrinkle to fold transition dominant in few layer numbers whereas higher layer numbers maintain their wrinkle like curvature.[4,10]

## S4. Layer dependent surface potential in 1-3L WS$_2$

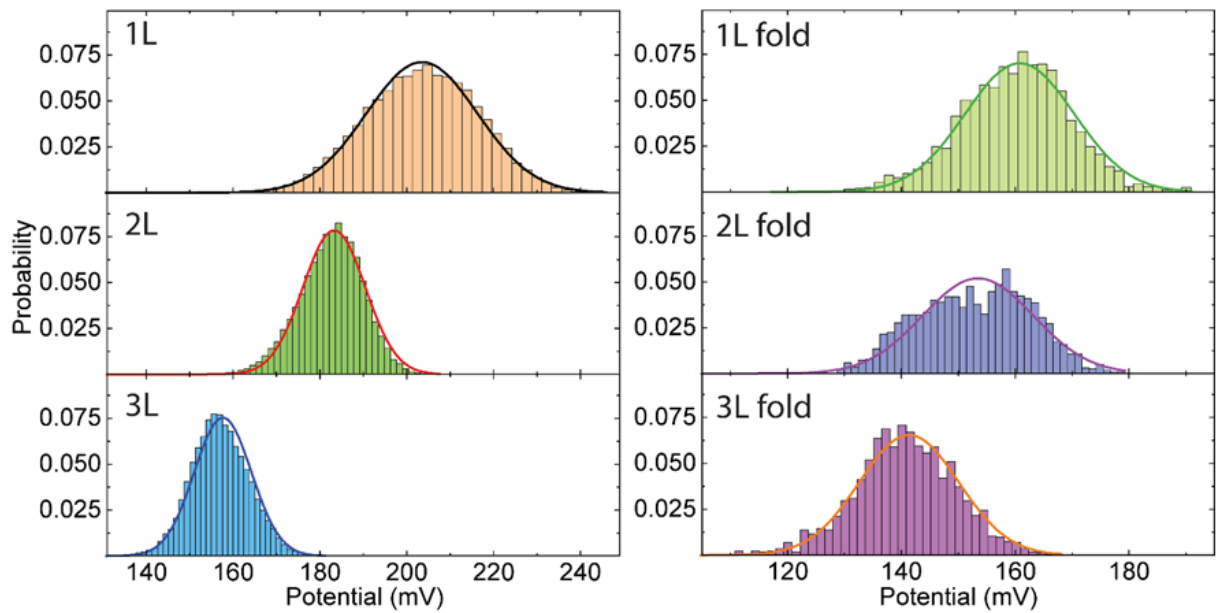

**Figure S3.** Histograms of the surface potential values collected at the 1-3L flat and folded regions

## S5. Dependence of photocurrent on bias voltage in strained WS$_2$

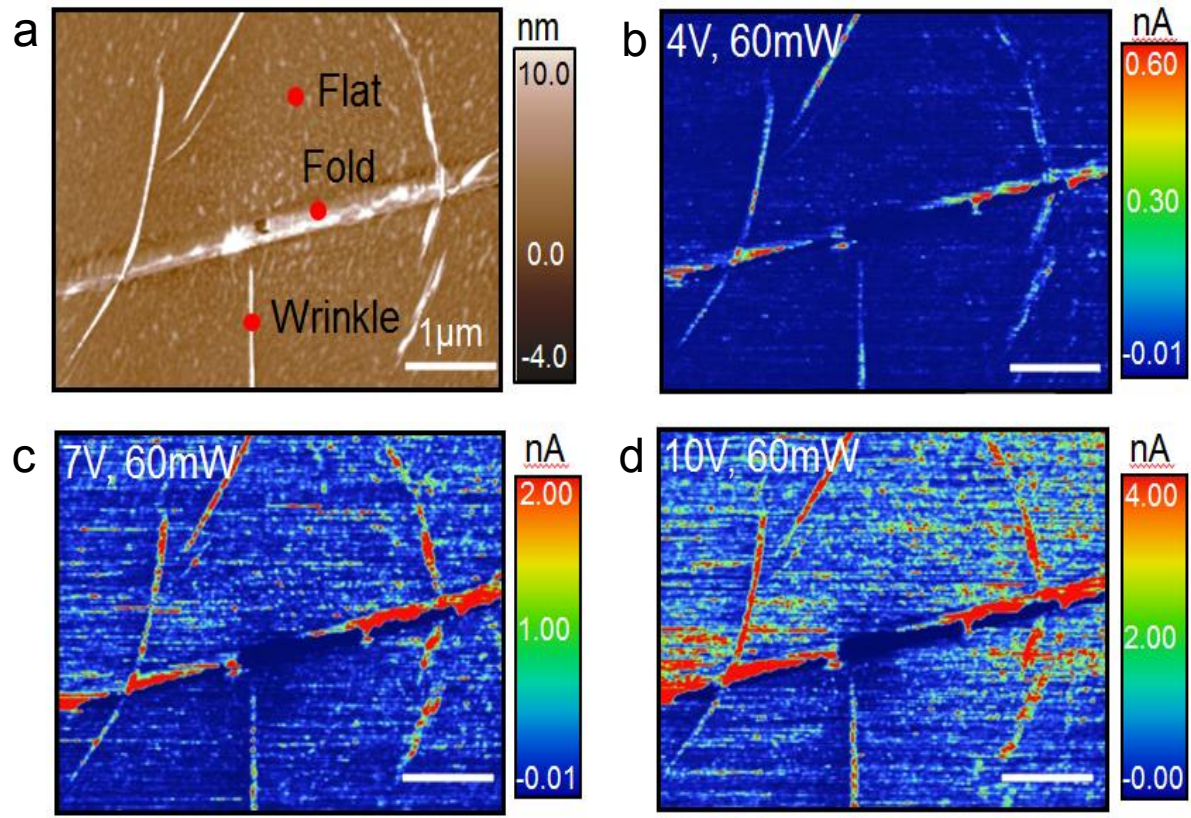

**Figure S4 | Dependence of photocurrent on bias voltage in 1L WS$_2$.** (a) Topographical image of the sample using atomic force microscopy showing flat, fold and wrinkle regions. (b)-(d) PCAFM based photocurrent maps of the sample in (a) at forward bias voltages of (b) 4 V, (c) 7 V, and (d) 10 V, scale bars in figure b-d represent 1μm.

## S6. Time-dependent photocurrent characteristics

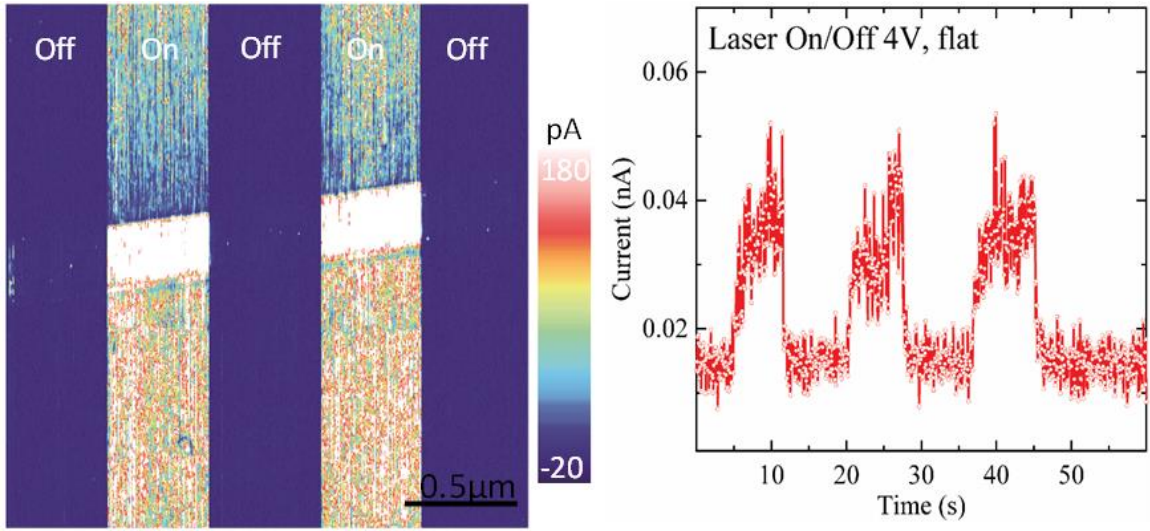

**Figure S5. Time-dependent photocurrent characteristics obtained under bias of 4 V with periodically switching on/off incident laser.**

## S7. Dependence of surface potential on laser power

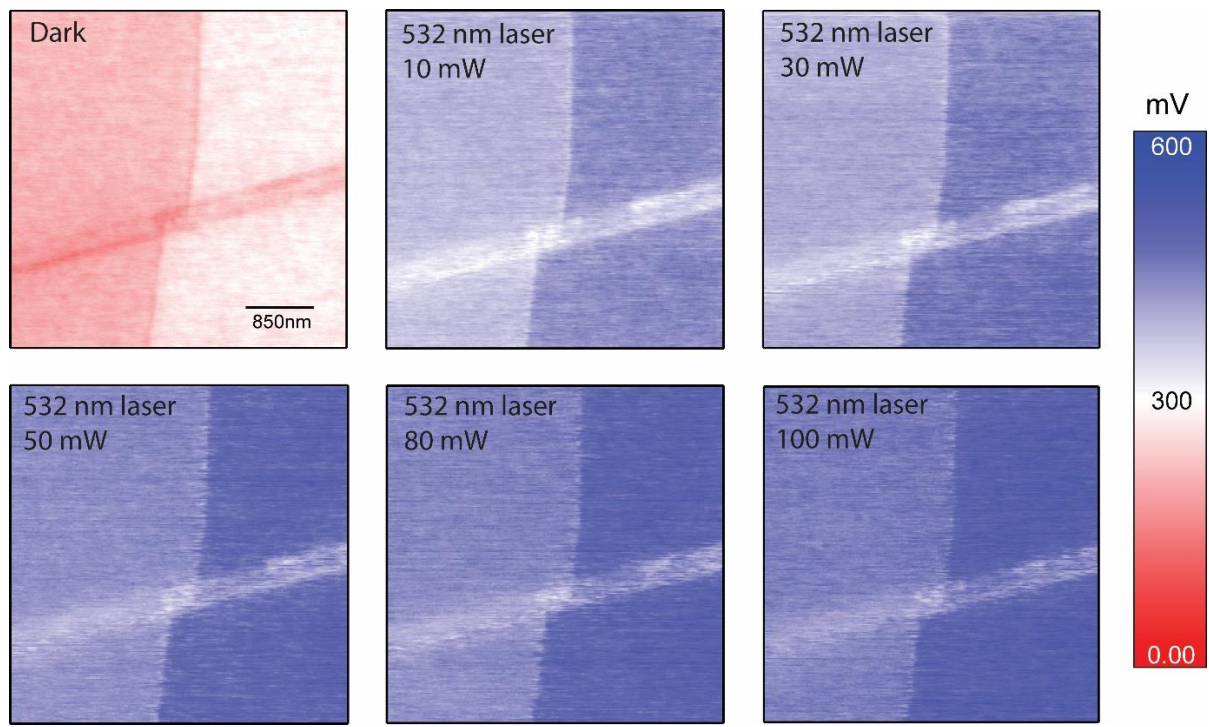

**Figure S6 | Dependence of surface potential on laser power.** The surface potential mapping on both 1L and 3L flat and wrinkled regions under different incident laser power.